\documentclass[%
 reprint,
 superscriptaddress,
nofootinbib,
 amsmath,amssymb,
 aps,
floatfix,
prd
]{revtex4-2}

\usepackage{graphicx}
\graphicspath{{Figures/}}
\usepackage{dcolumn}
\usepackage{bm}
\usepackage{hyperref}
\usepackage{cleveref}
\usepackage[separate-uncertainty=true]{siunitx}
\usepackage{caption}
\usepackage{subcaption}
\usepackage{placeins}
\usepackage{amsmath}

\begin{document}


\title{High-Performance Imaging in a Dilution Refrigerator} 

\author{Timo Eikelmann}
\altaffiliation{these authors contributed equally
}%
\affiliation{
 Institute for Quantum Physics and Center for Optical Quantum Technologies, University of Hamburg,
Luruper Chaussee 149, 22761 Hamburg, Germany
}%

\author{Mara Brinkmann}
\altaffiliation{these authors contributed equally
}%
\affiliation{
 Institute for Quantum Physics and Center for Optical Quantum Technologies, University of Hamburg,
Luruper Chaussee 149, 22761 Hamburg, Germany
}%

\author{Leonie Eggers}
\altaffiliation{these authors contributed equally
}%
\affiliation{
 Institute for Quantum Physics and Center for Optical Quantum Technologies, University of Hamburg,
Luruper Chaussee 149, 22761 Hamburg, Germany
}%
\affiliation{
 The Hamburg Centre for Ultrafast Imaging, Hamburg, Germany
}%
 
\author{Tuncay Ulas}
\affiliation{
 Institute for Quantum Physics and Center for Optical Quantum Technologies, University of Hamburg,
Luruper Chaussee 149, 22761 Hamburg, Germany
}%
 
\author{Donika Imeri}
 \affiliation{
 Institute for Quantum Physics and Center for Optical Quantum Technologies, University of Hamburg,
Luruper Chaussee 149, 22761 Hamburg, Germany
}%
\affiliation{
 The Hamburg Centre for Ultrafast Imaging, Hamburg, Germany
}%
\author{Konstantin Beck}
\affiliation{
 Institute for Quantum Physics and Center for Optical Quantum Technologies, University of Hamburg,
Luruper Chaussee 149, 22761 Hamburg, Germany
}%

\author{Lasse Jens Irrgang}
\affiliation{
 Institute for Quantum Physics and Center for Optical Quantum Technologies, University of Hamburg,
Luruper Chaussee 149, 22761 Hamburg, Germany
}%

\author{Sunil Kumar Mahato}
 \affiliation{
 Institute for Quantum Physics and Center for Optical Quantum Technologies, University of Hamburg,
Luruper Chaussee 149, 22761 Hamburg, Germany
}%
\affiliation{
 The Hamburg Centre for Ultrafast Imaging, Hamburg, Germany
}%

\author{Rikhav Shah}
\affiliation{
 Institute for Quantum Physics and Center for Optical Quantum Technologies, University of Hamburg,
Luruper Chaussee 149, 22761 Hamburg, Germany
}%

\author{Ralf Riedinger}
\email{ralf.riedinger@uni-hamburg.de}
\affiliation{
 Institute for Quantum Physics and Center for Optical Quantum Technologies, University of Hamburg,
Luruper Chaussee 149, 22761 Hamburg, Germany
}%
\affiliation{
 The Hamburg Centre for Ultrafast Imaging, Hamburg, Germany
}%

\date{\today}

\begin{abstract}
		\noindent 
        Nanophotonic light-matter interfaces hold great promise for quantum technologies. Enhancing local electromagnetic fields, they enable highly efficient detectors, can help realize optically connected processors, or serve as quantum repeaters. In-situ fiber-coupling at sub-Kelvin temperatures, as required for test and development of new devices, proves challenging as suitable cryogenic microscopes are not readily available. 
        Here, we report on a robust and versatile confocal imaging system integrated in a dilution refrigerator, enabling high-resolution visualization of nanophotonic structures on a transparent diamond substrate. Our imaging system achieves a resolution of $\SI{1.1}{\micro\meter}$ and a field-of-view of \SI{2.5}{\milli\metre}. The system requires no movable parts at cryogenic temperatures and features a large working distance, thereby allowing optical and microwave probe access, as well as direct anchoring of temperature sensitive samples to a cold finger, needed for applications with high thermal load.  
        This system will facilitate the development of scalable, integrated quantum optics technology, as required for research on large scale quantum networks. 
\end{abstract}

\maketitle

\section{Introduction}

Cryogenic microscopy is employed across various research fields, usually to characterize delicate samples, for in-situ alignment of probes, or collection of photons from fluorescent sources. 
One recent application field is quantum technology, often requiring cryogenically cooled samples, for example to induce superconductivity \cite{mirhosseini_superconducting_2020}, or to suppress thermal phonons \cite{Sukachev2017, Bhaskar2020} and anomalous electric surface noise \cite{Brownnutt2015}. 
State-of-the-art performance typically requires stringent control of the device environment, including, among others, temperature, vacuum, and magnetic fields. 
For example, silicon vacancy color centers in diamond (SiV), integrated in monolithic nanophotonic resonators---a promising platform for large scale quantum networks \cite{dawar2025quantuminternetresourceestimation}---require temperatures around \SI{100}{\milli\kelvin}, a bias field of \SI{0.5}{\tesla}, and ultra-high vacuum \cite{jahnke2015electron, Bhaskar2020, Stas2022}. Before permanent bonding of tapered fibers to an on-chip waveguide for highly efficient adiabatic mode transfer \cite{SIV310.1063/5.0170324, PhysRevApplied.8.024026, groblacher_highly_2013}, devices need to be characterized and scanned for suitable resonances and color centers. Coupling of these delicate sub-micron diameter fibers and waveguides below \SI{100}{\milli\kelvin} requires a dedicated photonic probe station. 

As standard microscopes are designed for ambient conditions and feature magnetic materials \cite{publicover_effects_1999}, such as steel springs, they are not compatible with, and require modifications to be used for quantum technology. 
There are two main approaches, which are also commercially available:

For wide-field imaging, a room-temperature long-working-distance objective can be used, placing a sufficient number of thermal and vacuum shields in-between the sample and the objective. This way, objectives and translation stages are kept at ambient conditions. This yields a relatively high imaging performance, at the cost of numerical aperture (NA), and a substantial reduction in effective working distance between the last window and the sample. As space constraints limit the number of thermal shield windows, resulting in substantial residual thermal load on the sample and cold finger of the cryostat, this approach is typically employed only down to liquid helium temperatures ($\sim\SI{4}{\kelvin}$). Examples include ion traps \cite{Pagano_2019,Miossec2022} as well as electronic and photonic probe stations. 

When higher collection yield or image resolution are required, the objective can be integrated in the cryogenic environment, allowing for shorter working distance by avoiding thermal shields altogether, resulting in higher NA. Here, specialized objectives are employed. For example, for correlative light and electron microscopy (CLEM) in structural biology \cite{pierson_recent_2024}, the relatively high working temperature of \SI{77}{\kelvin}  allows for partial cooling of the objective and positioning of the entire cold stage. At lower temperatures, as required for quantum dots and some color centers \cite{Sukachev2017, Senkalla2024, rinner_erbium_2023, PhysRevLett.124.023602}, the entire objective and nanopositioners are cooled, too. The nanopositioners are fragile and have limited dynamic force. This, in addition to the weight of the objective, creates limited thermal conduction between cold plate and sample (i.e.\ cross-section of thermal links), in particular when vapor cooling is not possible. 

Neither approach works well for sub-Kelvin optoelectronic probe stations, used, for example, for nanophotonic light-matter interfaces based on optomechanics or color centers: they require large working distances to accommodate probes, are sensitive to movement of the objective (vibration-induced), and require excellent thermalization of the sample under high vacuum condition. 

One simple approach is to replace the heavy long-working-distance objective of a wide-field microscope by an aspheric lens on nanopositioners \cite{riedinger_non-classical_2016, Stas2022}. As these feature no chromatic or plane-field corrections, monochromatic illumination is required and the effective field of view and image quality is substantially limited. Alternatively, a single achromat can be used. This improves chromatic and plane field performance, but the increased spherical aberrations limit the image quality, such that nanophotonic structures can barely be resolved \cite{MacDonald2015}. A parabolic mirror results in the opposite trade off for field planarity and spherical aberrations, but is notoriously hard to align while providing only a limited aberration corrected field-of-view \cite{Miossec2022}. 
It is possible to avoid imaging completely, and rely on other feedback signals, such as optical back reflection in the fibers \cite{ren_two-dimensional_2020, riedinger_remote_2018}, however, this severely limits alignment speed and setup versatility. 

Here, we report on an alternative approach to imaging of a nanophotonic probe station in a dilution refrigerator, keeping both sample and objective fixed. 
Focal shifts occurring during the cool-down process of the cryostat can be pre-compensated at ambient conditions, and corrected after cool-down by repositioning the external tube lens.
Using a non-magnetic, high-resolution wide-field objective, we achieve real-time visualization of nanophotonic structures over a large sample area, with the sample directly attached to the cold finger. A large working distance provides ample space for fiber-optic and coaxial probes. 
Employing confocal illumination by scanning of a white light laser, the system achieves an image quality comparable to room-temperature probe stations. 

In the following sections, we first describe the microscope setup, discuss design considerations and expected performance, and present experimental results. Finally, we provide a summary and a brief outlook. 


\section{Setup}
\label{sec:setup}
\begin{figure}[ht] 
    \includegraphics[width=0.56\linewidth]{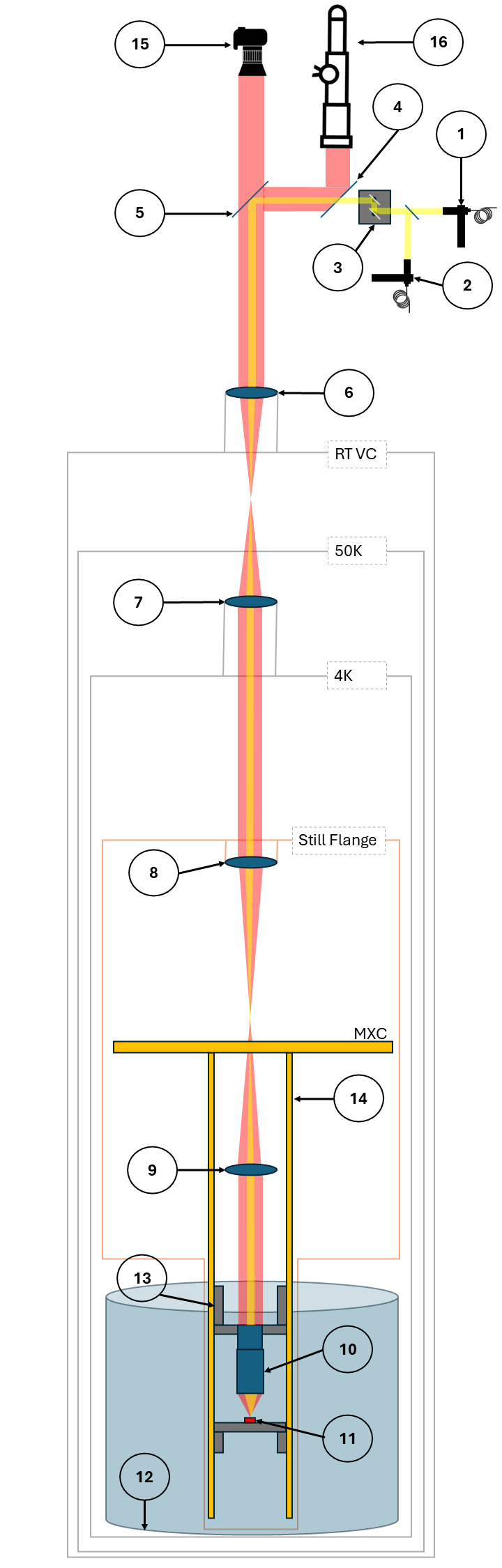}
    \caption{Schematic of the dilution refrigerator with all imaging components. The incoming beam is represented in yellow and the reflected light in red. For details see text~\ref{sec:setup}.}
    \label{fig:cryo_powerPoint}
\end{figure}

The main part of the microscope is an $8f$ confocal imaging system with a commercial $10\times$ infinity-corrected objective lens (see Fig.~\ref{fig:cryo_powerPoint}). The objective is located in the bore of a superconducting vector magnet on the mixing chamber stage of a dilution refrigerator. Four relay lenses (the $8f$ system configured as two subsequent $4f$ lens systems, each lens positioned two focal lengths $f$ after the other, acting as a one-to-one imaging system) transfer the light collected by the objective across the different temperature stages of the cryostat. At a distance of approximately twice the focal length of the last relay lens, which also serves as the vacuum window, the collected light recovers the spatial and angular properties it had at the back aperture of the objective. 
This virtual objective back aperture, the $8f$-point, is imaged by tube lenses, and illuminated by scanning galvanometer mirrors, to provide confocal illumination. Positioning of the tube lens, respectively camera, allows for focusing the region of interest (ROI) within a limited range. 

In the following, we provide a detailed description of the setup illustrated in Figure~\ref{fig:cryo_powerPoint}. 
Further information on the used components can be found in the Appendix~\ref{sec:appendix}. 

The system uses a collimated white light laser (Fig.~\ref{fig:cryo_powerPoint}-1) or monochromatic green laser (\ref{fig:cryo_powerPoint}-2) for illumination. These beams are overlapped on a non-polarizing beam splitter (BS), and reflected by two galvanometer mirrors (\ref{fig:cryo_powerPoint}-3), allowing a rapid scanning over an area of about $\SI{1}{\mm^2}$ on the sample. 
The light then passes two non-polarizing 50/50 splitters (\ref{fig:cryo_powerPoint}-4,5), which are used to direct returning light to two cameras.
Thereafter, the light goes through 
the flange lens (\ref{fig:cryo_powerPoint}-6), an achromat with $f=\SI{250}{\milli\metre}$ focal length, entering the dilution refrigerator. It simultaneously serves as the last relay lens, and optical vacuum port to the refrigerator. Due to the curved surface, ghosting as well as back-reflection of the illumination lasers into the camera are avoided. The silica-based lenses absorb the dominant parts of the thermal radiation of objects from 100 to \SI{500}{\kelvin} (wavelengths $\SI{60}{\micro\meter}$ to $\SI{5}{\micro\meter}$, covering $>75\%$ of the black body emission). Hence, excess thermal radiation from hot sources, such as camera sensors, is prevented from reaching the lower stages of the dilution refrigerator. 
Inside the dilution refrigerator, three more lenses (\ref{fig:cryo_powerPoint}-7,8,9) are placed on different temperature stages, with each distance matching the sum of both focal lengths (\ref{fig:cryo_powerPoint}-6 to 7: $\SI{500}{mm}$, 7 to 8:$\SI{450}{mm}$, 8 to 9: $\SI{400}{mm}$), resulting in the $8f$ configuration. Achromatic lenses are chosen due to their wavelength-independent focal length, which is essential for a white light laser source, plane field correction, and absorption of mid-infrared thermal radiation. The different focal lengths of the two $4f$ relays (\ref{fig:cryo_powerPoint}-6,7: $f_6=\SI{250}{mm}$, 8,9: $f_8=\SI{200}{mm}$) are chosen such that the lenses are located in convenient heights relative to the cold stages, simplifying the mounting. 

The objective (\ref{fig:cryo_powerPoint}-10) has a focal length $f_\text{o}=\SI{20}{mm}$, and is positioned $\SI{200}{mm}$ after the last achromat (\ref{fig:cryo_powerPoint}-9). It focuses the laser onto the sample (\ref{fig:cryo_powerPoint}-11) in the experimental volume hosted below the mixing chamber (MXC) and surrounded by the vector magnet (\ref{fig:cryo_powerPoint}-12). The objective is mounted on an objective holder (\ref{fig:cryo_powerPoint}-13), which can be adjusted in height at ambient conditions using two removable micrometer screws. The holder, sample mount and the support structure (\ref{fig:cryo_powerPoint}-14), on which they are mounted, are all made of gold-plated copper, such that their relative thermal contractions are compensated.\\
Light scattered and reflected from the sample (red) retraces its path up to the beamsplitter (\ref{fig:cryo_powerPoint}-5) above the flange lens (\ref{fig:cryo_powerPoint}-6). Here, part of the beam is transmitted to the digital tele-zoom camera (\ref{fig:cryo_powerPoint}-15, $\SI{55}{\milli\meter}<f_\text{z}<\SI{200}{\milli\meter}$) for wide-field imaging, while the reflected part follows the path towards the other beamsplitter (\ref{fig:cryo_powerPoint}-4) and into the telescope (\ref{fig:cryo_powerPoint}-16, $f_\text{t}=\SI{520}{\mm}$) for high-resolution imaging. 
These two telescope objectives (\ref{fig:cryo_powerPoint}-15, 16) serve as motorized tube lenses, to facilitate the external focusing at cryogenic temperatures. 
The difference in focal length provides a choice of magnification between sample and camera, nominally varying between $3< f_\text{z}/f_\text{o} < 10$ for the wide-field camera, to maximize the field of view. The \SI{30}{\milli\meter} camera sensor achieves a usable field diagonal of $\sim \SI{2.5}{\milli\meter}$, limited by the diameter of the line-of-sight port of the cryostat. 

The magnification of the high-resolution setup is $M=f_\text{t}/f_\text{o}=26$, which is sufficiently large, such that the \SI{2.3}{\micro\meter} camera pixels nominally correspond to an image-size of \SI{90}{\nano\meter}, far smaller than the optical resolution limit. 
The camera is a cooled CMOS chip without an infrared filter, allowing for imaging in low-light conditions, and detecting light at the SiV resonance wavelength ($\lambda=\SI{737}{\nano\meter}$).
To select the ROI with the high-magnification setup, the beamsplitter (\ref{fig:cryo_powerPoint}-4) is equipped with two motorized micrometer screws, allowing for minute adjustment of the angle close to the $8f$ virtual objective back aperture. 

Both illumination laser and high magnification camera can be equipped with optical band pass filters, when suitable for the application. This is useful, for example, when simultaneously imaging nanophotonic samples and an illuminated tapered fiber during device coupling, see below.

\section{Design considerations and System response}

The design goal is to achieve high resolution hyperspectral imaging of low-contrast structures under minimal heat load employing fixed cryogenic optics. 
We first discuss the design of the $8f$ confocal setup, primarily addressing the first two points. Thereafter, we discuss the pre- and post-compensation of focal shifts during the cooling of the cryostat. 

Confocal illumination has been chosen for imaging under minimal light conditions. By adjusting the scan range of the galvanometer, which is positioned in the focus of the last relay lens (\ref{fig:cryo_powerPoint}-6), the illumination can be restricted to the ROI. It further is beneficial for the resolution, by increasing the NA of the condenser compared to a diffuse light source in the cryostat. As nanophotonic structures on transparent substrates such as diamond feature very low visible contrast compared to the solder or thermal adhesive on the opposite side, scanning confocal illumination drastically improves the imaging quality by reducing depth-based artifacts.
It was found in a previous iteration of the setup that spherical aberrations and field curvature of the relay achromats have negligible  influence on the resolution in the central part of the image~\cite{Ulas_thesis}. 

The system is in principle suitable for confocal fluorescence imaging of color-centers in diamond \cite{AGruber1997}, albeit the low NA of $0.45$ of the objective requires high defect densities or resonant enhancement. The green diode laser ($\lambda\approx \SI{520}{nm}$) excites SiV color-centers and a high extinction band-pass in front of the camera blocks the returning green light while transmitting the fluorescence at $\lambda \approx \SI{737}{\nano\meter}$.

\begin{figure*}[t]
\includegraphics[width=\linewidth]{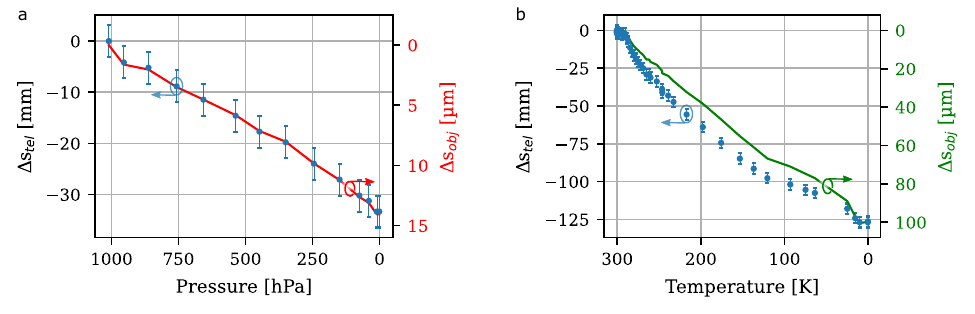}
\caption{Change of focal plane due to (a) pressure and (b) temperature changes in the dilution refrigerator. Left axis ($\Delta s_\text{tel}$): Measured change of image plane (blue), error bars indicate depth of focus $\sim \SI{6}{\milli\meter}$. Right axis ($\Delta s_\text{obj}$): Extrapolated change of distance between sample and objective-side principle plane (red, green). Circles indicate applicable vertical axis.}
	\label{fig:pump_shift}
\end{figure*}

The imaging system is installed in a dilution refrigerator, and operated at low temperatures and pressures for which the employed optical components are not rated. 
Hence, the quality and utility of the setup needs to be characterized during cool-down and at base-temperature of the 
dilution refrigerator.
During the cool-down process, the system is first evacuated to a pressure of $\sim10^{-3}\,\si{\hecto\pascal}$, and subsequently cooled by a pulse-tube and $^3$He/$^4$He mixture until a final temperature of $\sim 10\,\si{\milli\kelvin}$ and a pressure of $\sim 10^{-6}\,\si{\hecto\pascal}$ is reached. The pressure in the sample space is substantially lower due to cryo pumping, but cannot be measured by room-temperature gauges. 

Due to cryo-pumping, the objective does not out-gas, and thus does not negatively affect the final pressure near the sample. To avoid large strain as well as excessive nitrogen ice formation within the objective, it is advisable to use an objective without hermetically sealed internal volumes. While vendors do not usually provide this information, X-ray tomography of the objective indicates that the objective used here can be evacuated. It was further found that drilling evacuation holes into the objective body was not necessary, as virtual leaks of air pockets inside the objective do not limit the evacuation speed of the system. 
We further did not observe a degradation of the anti-reflection coatings or optical cement during cool-downs.

The expected effect on the imaging system during cool-down stems from thermal contractions, changes in refractive index, and pressure-related shifts of lens positions. 
While all these effects on the lens system can be calculated in theory, it is not possible to get accurate predictions without precise knowledge of the inner construction of microscope and telescope objectives, or a black-box simulation description, which was not provided by the vendors. Hence, we proceed with initial estimates, approximating them as simple lenses with their respective focal lengths. 

While the $8f$ system is fairly robust against small variations, the optical distance between sample and objective has the largest effect.
A shift of the ROI away from the objective ($s_\text{o}\rightarrow s_\text{o}+\Delta s_\text{obj}$), results in a shift of the tube-lens focal plane towards it ($s_\text{t}\rightarrow s_\text{t}+\Delta s_\text{tel}$) by 
\begin{equation}
    \Delta s_\text{tel} \approx -M^2 \Delta s_\text{obj},
\end{equation}
with the magnification $M=s_\text{t}/s_\text{o}$ corresponding to the ratio of distance of the focal planes from principle planes of the lens system on the objective side ($s_\text{o}$), respectively on the ocular side ($s_\text{t}$, see \ref{si:focal}).
At the nominal working point, this corresponds to $f_\text{z}/f_\text{o}$
for the zoom lens (\ref{fig:cryo_powerPoint}-15), respectively $f_\text{t}/f_\text{o}$ for the telescope (\ref{fig:cryo_powerPoint}-16). 
As the magnification changes substantially with shifting focal planes, the objective-side focal shift can be computed by integrating ocular-side focal shifts over the measured magnification. In this setup a total shift $\Delta s_\text{obj,max}=\SI{0.15}{\milli\meter}$, corresponding to $\Delta s_\text{tel,max}=\SI{0.25}{\meter}$ can be compensated. 
In the following paragraphs, we estimate the focal shifts due to evacuation and thermal contraction, comparing them to the system limit.

As the refractive index of air decreases with pressure, the ratio of the refractive indices between the lens and the air are reduced, leading to a shorter focal length
\begin{equation}\label{eq:f_change_evacuation}
    f' = f\left(1 + \delta n_\text{air} \left(1+\frac{1}{n_\text{lens}-1}\right)\right), 
\end{equation}
with the refractive index shift $\delta n_\text{air} = 3\cdot 10^{-7} \delta p/\text{hPa}$ depending on the differential pressure $\delta p$, assuming a thin lens. For our objective, this corresponds to a shift $\SI{14}{\micro\meter}<\Delta s_\text{obj,air}<\SI{20}{\micro\meter}$ (see \ref{si:focal}). This is within the correctable limit $\Delta s_\text{obj,max}$. We note that the negative change in refractive index of air leads to a shortening of focal length and thus an effective shift of the sample away from the objective. If air is trapped inside the objective, the plano-convex shape of the lens would mean that there is virtually no effect on the focal length, and the internal pressure would potentially shift the front lens towards the sample, i.e.\ a negative shift $\Delta s_\text{obj}<0$ would occur.  In Fig.~\ref{fig:pump_shift}a, we observe a shift of the focal plane towards the tube lens, consistent with the objective being evacuated. This shift corresponds to a reduction of the objective's focal length by $\sim\SI{14}{\micro\meter}$, using the experimentally measured magnification of $M=48$ during evacuation (see Fig.~\ref{fig:pump_shift}a and \ref{si:magnification}). This is consistent with the high-index primary lens ($n_\text{lens}\sim1.7$) of the objective. 

With decreasing temperatures, materials in the dilution refrigerator contract, altering the distances between optical components. 
To minimize the effect on the 
distance between the sample and the objective lens, the system can be constructed from material combinations, such that the thermal contractions are compensated. For the objective used here, which we assume is constructed predominantly from aluminum and brass, this can be achieved using copper side-rails (\ref{fig:cryo_powerPoint}-14) and objective mount (\ref{fig:cryo_powerPoint}-13) without an additional spacer, yielding an expected shift of the sample to objective of $-\SI{16}{\micro\meter}<\Delta s_\text{obj, th}<\SI{6}{\micro\meter}$.
As the material composition is unknown, this estimate carries a large systematic uncertainty: other metals, or thermoplastic polymer spacers could extend this range to 
$-\SI{0.2}{\milli\meter}<\Delta s_\text{obj, th}<\SI{0.6}{\milli\meter}$ (for titanium, respectively PMMA, as lens tube material in the objective). 

Further, the thermal contraction of lens material itself changes its radius of curvature, resulting in a reduction of focal length. Assuming typical silica-based optical glasses, the relative cumulative contraction should not exceed $\Delta L/L < 0.15\%$, corresponding to a focal shift $\Delta s_\text{obj,rad}<\SI{30}{\micro\meter}$ for the objective. 

\begin{figure}[t] 
	\includegraphics[width=0.75\linewidth]{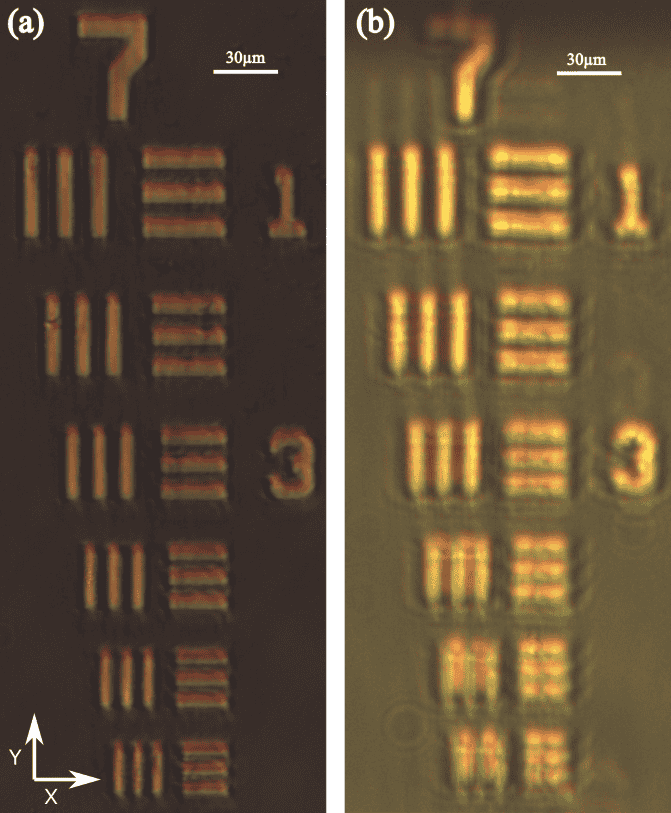}
	\caption{Exemplary images of group 7 of the USAF-1951 test target taken at room temperature (a) and cryogenic temperatures (b) with the high-magnification tube lens.}
	\label{fig:cold_and_warm_Usaf}
\end{figure}

While the dependence of refractive index on temperature typically dominates the behavior of optical elements near room temperature, the thermo-optic coefficient reduces, or changes its sign for numerous silica-based glasses, at cryogenic temperatures, leading to non-linear temperature dependence and little overall effect \cite{leviton_temperature-dependent_2006, frey_cryogenic_2007}. For a single lens experiencing a refractive index change $\delta n_\text{lens}$, the focal length decreases to 
\begin{equation}\label{eq:f_change_therm}
    f' = f \left(1 - \frac{\delta n_\text{lens}}{ n_\text{lens}-1}\right).
\end{equation}
Approximating the objective as a single lens, this would result in an effective focal shift $\Delta s_\text{obj,n}\approx f-f'$ in the range of $-\SI{64}{\micro\meter}<\Delta s_\text{obj,n}<\SI{8}{\micro\meter}$, for fused silica or BaLKN3, both chosen as examples with large cumulative shifts\cite{leviton_temperature-dependent_2006, frey_cryogenic_2007}. 

In summary, we expect an overall shift in the range of $-\SI{80}{\micro\meter}<\Delta s_\text{obj,p}<\SI{44}{\micro\meter}$. 
The largest systematic uncertainty is related to the thermal contraction of the objective due to the unknown material composition.
Note that the thermal contraction of spacers and lenses partially compensate for the relay lenses, such that the overall contribution is expected to be small. 

As shown in Figure~\ref{fig:pump_shift}b, the observed shift $\Delta s_\text{obj,obs}\sim \SI{0.10}{\milli\meter}$ is twice as large as estimated. Nevertheless, it is smaller than the uncompensated thermal contraction, and within the range $\Delta s_\text{obj,max}$ that can be compensated by the large magnification tube lens (\ref{fig:cryo_powerPoint}-16). As the shift of the focal plane is approximately linear in temperature, we attribute this to thermal contraction, rather than a thermo-optic effect. This is consistent with the use of a polymer spacer in the objective, which was also indicated by X-ray tomography. The focal shift during evacuation and cool-down is consistent between thermal cycles, such that it can be pre-compensated at ambient conditions, aligning the objective such that the tube lens focal plane is located $\sim\SI{16}{\centi\meter}$ above the nominal back-focal plane of the telescope. Hence, after cool-down, the system approaches its nominally optimal configuration. 

\begin{figure}[t]
    \centering
    \includegraphics[width=\linewidth]{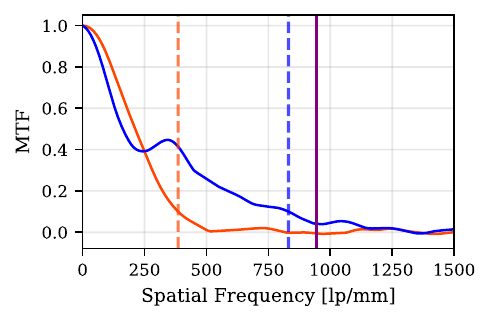}
    \caption{MTF analysis for the Y-axes of the images in Fig.~\ref{fig:cold_and_warm_Usaf}, at ambient conditions (blue, Fig.~\ref{fig:cold_and_warm_Usaf}a) and $T\approx\SI{70}{\milli\kelvin}$ (red, Fig.~\ref{fig:cold_and_warm_Usaf}b). The resolution is stated as the value where the background subtracted MTF drops to 10\% (dashed lines, b), yielding $\SI{831}{\frac{lp}{mm}}$ at room temperature and $\SI{385}{\frac{lp}{mm}}$ at cryogenic temperature in this example. The purple line indicates the estimated technical limit of the optical system ($\SI{945}{\frac{lp}{mm}}$).}
    \label{fig:MTF_combined}
\end{figure}

\begin{figure*}[t] 
	\includegraphics[width=\linewidth]{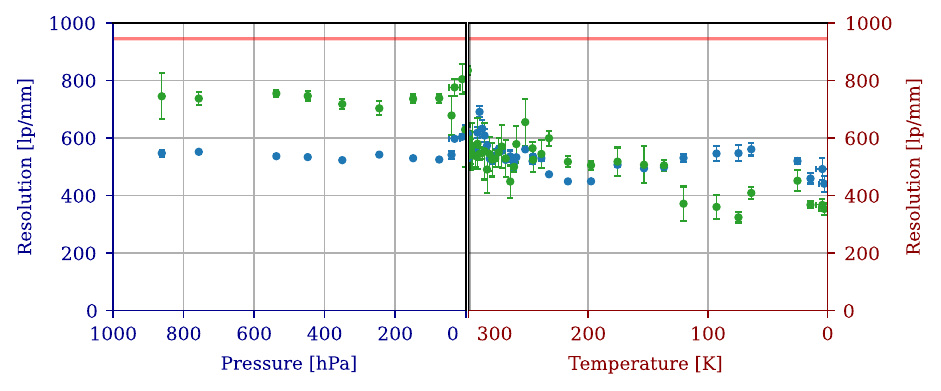}
	\caption{Change of resolution during evacuation (left) and cool-down (right). Resolution corresponds to average MTF10\% value of repeated optimization (blue: X-axis, green: Y-axis, see Fig.~\ref{fig:cold_and_warm_Usaf}). Error bars indicate statistical standard deviation from repeated optimization. 
    The red line indicates $945 \frac{lp}{mm}$, corresponding to 70\% of the nominal resolution of the objective. 
    The average resolution at temperatures below \SI{100}{\kelvin} is $443\pm78 \frac{lp}{mm}$. In the beginning of the cool-down process, a higher data collection rate was chosen.}
	\label{fig:res_pump}
\end{figure*}

Beyond the change in focal distance, the image quality can be further affected during cool-down by condensation and modification of the lens system. As discussed above, we do not expect trapped air in the objective, and we do not observe signs of nitrogen ice formation inside the objective. However, we expect aberrations due to the shift of lenses within the objective, reducing the achievable resolution of the system. These aberrations are characterized in the following section.

\section{Results and Discussion}

To determine the resolution of the imaging system, pictures of a USAF 1951 test pattern (see Fig.~\ref{fig:cold_and_warm_Usaf}) were taken and analyzed. For each set of rectangular triplets in the test pattern, an edge was analyzed to obtain its respective Modulation Transfer Function (MTF):
The edge spread function is first numerically calculated from the imaged edges, and then its spatial derivative is Fourier transformed, yielding the MTF (see Fig.~\ref{fig:MTF_combined}). 
The nominally achieved resolution (in line-pairs per mm, \si{\frac{lp}{mm}}) of a rectangle triplet is determined at the point where the MTF falls to a value 10\% of its maximum (henceforth denoted as MTF10\%). The mean value of the resolutions corresponding to MTF10\% across all lines in the image gives the resolution for the entire image, shown in Fig.~\ref{fig:res_pump}. The arbitrary 10\% limit is usually assumed to correspond to a level of resolution that can still be discerned by the human eye.

In Figure~\ref{fig:MTF_combined} the MTF for an image at room temperature (red line, corresponding to fig.~\ref{fig:cold_and_warm_Usaf}(a)) and for an image at \SI{70}{\milli\kelvin} (blue line, corresponding to fig.~\ref{fig:cold_and_warm_Usaf}(b)) with their respective MTF10\%, indicated through the dashed lines, are shown.
The limit of the physical resolution, given by the NA and the wavelength of the used light source, is indicated in purple and gives a maximum resolution $\SI{1455}{\frac{lp}{mm}}$ corresponding to the MTF10\%, for a wavelength of \SI{500}{\nano\meter} and NA of $0.45$. According to specifications, this is reduced to $\SI{1350}{\frac{lp}{mm}}$ for the objective,  and we expect a further reduction to $\sim\SI{945}{\frac{lp}{mm}}$ due to the typical performance of this type of tube lens (Fig.~\ref{fig:cryo_powerPoint}-16), all for on-axis image components. 

At room temperature, the high-magnification imaging system exhibits a resolution of $\sim\SI{800}{\frac{lp}{mm}}$, slightly less than the rough estimate above. This is to be expected, as each $4f$-system introduces aberrations, and the tube lens is operated away from its designed back focus in order to compensate for later thermal shifts. Furthermore, the resolution at cryogenic temperatures $T<\SI{100}{mK}$ is further reduced to $\sim\SI{400}{\frac{lp}{mm}}$, which is also expected, as the optimization of the system can only be done at room temperature with an open system, and additional aberrations ought to manifest as the focal lengths and position of lenses change. Overall, the resolution remains sufficient to distinguish neighboring nanophotonic structures.

We monitor the change in  MTF10\% resolution during the evacuation and cooling (see Fig.~\ref{fig:res_pump}). The focus plane of the high-magnification system was continuously adjusted to optimize the resolution in Y-direction (see Fig.~\ref{fig:cold_and_warm_Usaf}). The error bars indicated the standard deviation of MTF10\% values for different triplets in one image. 
We observe a substantial astigmatism, which changes with pressure and temperature, due to the ROI being off-center with respect to the optical axis, representative for the use as nanophotonic probe station. The objective is not fully plane-field corrected and typically has a small tilt with respect to the sample surface, such that neither both axes, nor the entire image can be perfectly in focus at all times. We observe that the X-axis resolution remains approximately the same during the evacuation and cool-down, while the Y-axis resolution changes substantially. A pronounced drop between the fully evacuated system and the start of the cool-down is related to the vibrations of the pulse tube. A slight further reduction in resolution upon cooling to cryogenic temperatures is likely related to a tilt of the objective focal plane along the Y-axis, also indicated by the larger variance in MTF10\% values. If the ROI is known before closing the cryostat, the astigmatism can be compensated manually by centering the sample and tilting the objective. Eliminating it completely requires substantial practice, however, and is not required for usage as nanophotonic probe station. 

The best MTF10\% resolution achieved during this cool-down below $\SI{100}{\kelvin}$ is $\SI{440\pm 80}{\frac{lp}{mm}}$, corresponding to a resolution of $\xi=\SI{1.1\pm0.2}{\micro\meter}$. It  stays stable below \SI{100}{\kelvin} down to the base temperature of the dilution refrigerator of approximately \SI{10}{\milli\kelvin} (see Fig.~\ref{fig:res_pump}).

\begin{figure}[t] 
	\includegraphics[width=\linewidth]{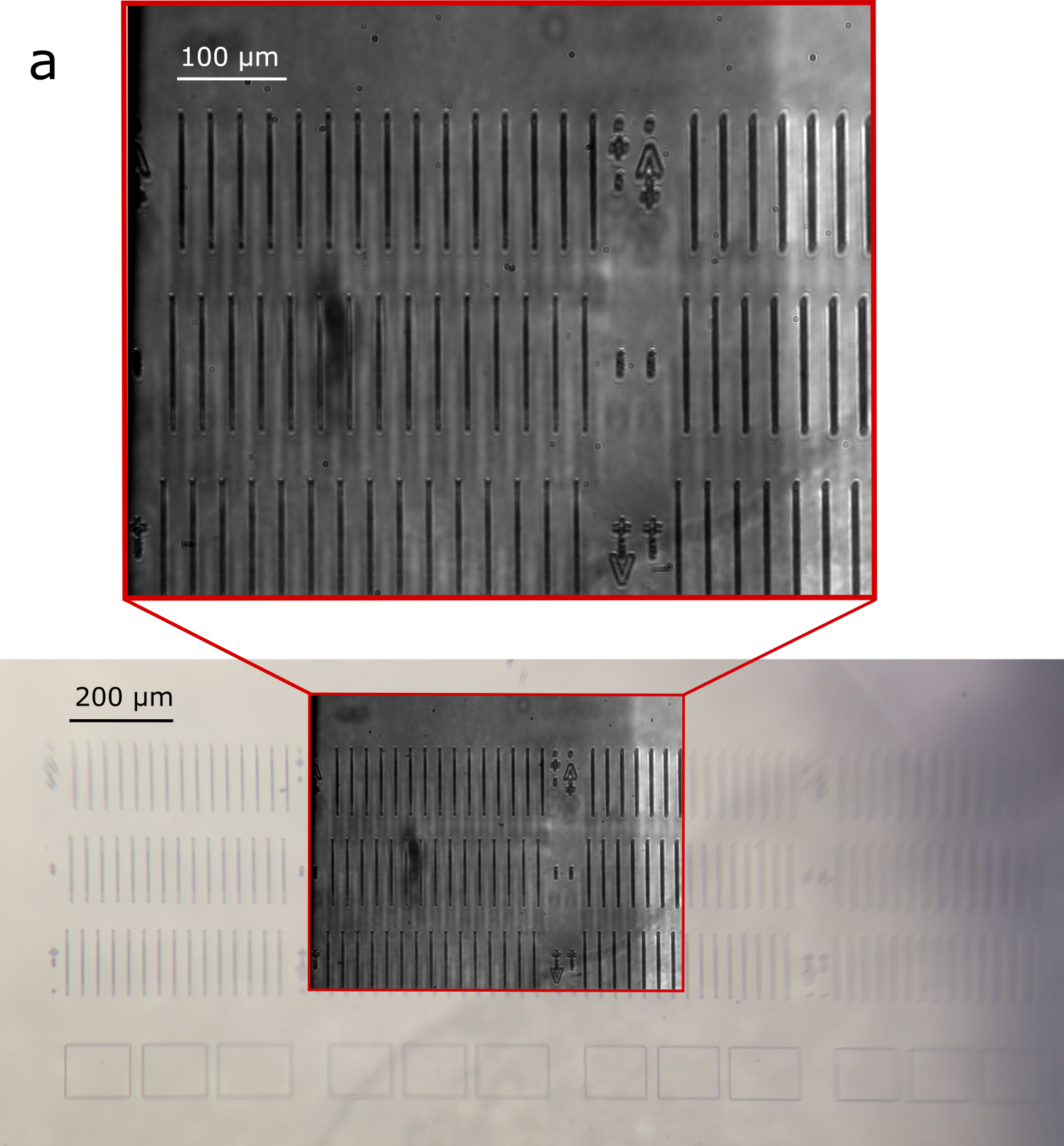}
    \includegraphics[width=\linewidth]{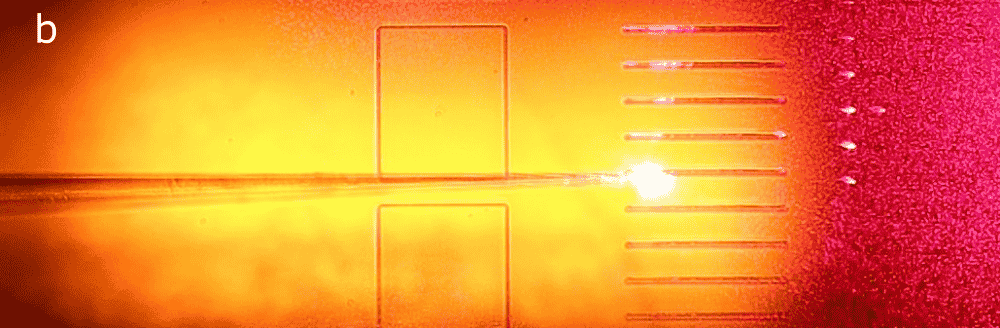}
	\caption{Image of nanophotonic structures on a diamond chip. (a) Wide-field image captured in the dilution refrigerator at ambient conditions. The vertical lines are waveguides with optical resonators. Inset: high-magnification setup obtained with objective and sample anchored to the mixing chamber at a temperature of approximately $\SI{10}{\milli\kelvin}$.
    Sample and objective are in vacuum. In the areas far from the center of the images, spherical aberrations are visible, but have been deemed acceptable for further use of the imaging system. (b) Wide-field image during the coupling procedure of a tapered fiber to a waveguide, under confocal illumination (yellow) of a small ROI, with a red laser (wavelength \SI{650}{\nano\meter}, power \SI{6}{\micro\watt}) coupled to the fiber for visualization purposes.}
	\label{fig:nanophotonic_structures}
\end{figure}

The resolution of the imaging system can be affected by numerous internal and external factors. While we do not have access to the system when evacuated, or cooled, the system response indicates some limiting features: 

First, the difference between expected and observed focal shift indicates that the primary lens of the objective shifts in position more strongly than the others, by approximately \SI{0.1}{mm}, which is expected to result in increased aberrations. We do not observe signs of substantial strain in either the objective or relay lenses. 
Second, the $8f$-relay lenses are fixed in position, but not manually optimized for image performance. This results in a robust system, but can introduce additional aberrations, which can be further aggravated during evacuation and cool-down: the first lens (Fig.\ref{fig:cryo_powerPoint}-6) rests on an O-ring that is compressed during vacuum pumping, changing the positioning of the lens. The relay lenses thus deviate from an ideal one-to-one imaging system, changing beam divergence and overall magnification. Third, the inner diameter of the line-of-sight port is $\SI{40}{\mm}$, cutting off higher spatial frequencies in the Fourier plane for off-axis components of the image plane, in particular for a diverging beam. This visually limits the resolution of the wide-field tube lens (Fig.\ref{fig:cryo_powerPoint}-15, see also Fig. \ref{fig:nanophotonic_structures}, bottom). Fourth, the reduced magnification increases the effective f-number of the tube lens, potentially increasing the effect of aberrations, despite approaching its nominal back focus. 

The imaging system can be further influenced through external factors: at base temperature, the imaging system was exposed to a strong magnetic field up to \SI{2}{\tesla} along the $z$-axis and \SI{0.5}{\tesla} along the $x$- and $y$-axes. No effect on the image quality was observed, consistent with the assumption of a non-magnetic objective. Vibrations from the pulse tube cryo-cooler can also influence the image, in particular for a cryostat with parallel support structures. 
This effect was less pronounced when a superconducting vector magnet was installed, due to its additional weight of \SI{24}{\kilo\gram} anchoring the microscope. We see that only one axis is affected substantially, indicating that stiffening the support structures can suppress these issues. It would further be possible to synchronize the scanning galvanometer with the pulse tube cycle to reduce its impact. 

The system was then employed to image nanophotonic structures on a diamond chip, as seen in Fig.~\ref{fig:nanophotonic_structures}. Here, the wide-field image (Fig.~\ref{fig:nanophotonic_structures}a, bottom) was captured at room temperature under ambient light, while the smaller inset was taken with the telescope camera (Fig.~\ref{fig:cryo_powerPoint}-16) under cryogenic conditions at a temperature of \SI{10}{\milli\kelvin}, using white light confocal illumination with a power up to \SI{20}{\micro\watt}. The variation in illumination is due to the fact that the galvanometer scan is not synchronized with the camera shutter. The objective holder is slightly tilted and the objective is non-plane corrected, which results in areas far from the center of the image appearing out of focus. This is mainly visible when using the auxiliary camera, because of the larger field-of-view. 

On all images in Fig.~\ref{fig:nanophotonic_structures}, the  waveguides, pictured as the dark lines (\SI{550}{\nano\meter} wide and \SI{125}{\micro\meter} long), and the alignment markers between the sets of waveguides are easily recognizable. The $\sim\SI{1.0}{\micro\meter}$ wide waveguide supports can be discerned with an optimized focus, consistent with the calculated resolution of \SI{1.1}{\micro\meter}.
The darker shade in the background is the silver-filled adhesive used to fix the diamond chip to the copper sample holder. Structures far from the center of the image are out of focus as a result of a tilted focal plane. The right side of the wide-field image (bottom) is darker and blurry due to clipping on the sides of the line-of-sight port used by the $8f$-relay, whereas the left side is substantially sharper, as it is close to the optical axis of the system.
The wide-field system features a full field of view of $\sim\SI{2.5}{\milli\meter}$ ($\sim\SI{2.6}{\milli\meter}$) in X (Y) direction, albeit not the entire FOV can be in focus simultaneously. Due to the smaller magnification, the wide-field imaging system can be used to rapidly shift the focal plane, as required for a careful approach with a tapered optical fiber (Fig.~\ref{fig:nanophotonic_structures}b).

\section{Summary and Outlook}
We report on the implementation of a confocally illuminated imaging-system in a dilution refrigerator, reaching a resolution of $\SI{440\pm 80}{\frac{lp}{mm}}$ below \SI{100}{\kelvin}, and a field of view of \SI{2.5}{\milli\meter} at temperatures below \SI{100}{\milli\kelvin}. The system requires no moving parts at cryogenic temperatures, and induces minimal heating, allowing for probe access and excellent thermalization of samples in ultra-high vacuum and at a mixing chamber temperature of $\sim\SI{10}{\milli\kelvin}$.
The changes in the resolution, the focal plane shift and the field-of-view due to the changes in pressure and temperature are characterized and discussed to aid the understanding of the behavior of optical systems at cryogenic temperatures.\\
The system is then employed to image a diamond sample with nanophotonic structures at approximately \SI{10}{\milli\kelvin}. 
In the future, the system will be suitable to monitor evanescent coupling of tapered optical fibers to nanophotonic structures on a diamond chip for quantum interfaces with high efficiency.
A high resolution microscope in a \SI{10}{\milli\kelvin} optical probe station with image quality similar to an equivalent system under ambient condition will facilitate the development of scalable, integrated quantum interfaces, to be used as network nodes in a large scale quantum internet testbed.

\vspace{18pt}
\begin{acknowledgments}
The authors want to thank Luca Graf for valuable discussions.
This research was supported by the Cluster of Excellence ‘Advanced Imaging
of Matter’ of the Deutsche Forschungsgemeinschaft (DFG)–EXC 2056–Project ID 390715994, Bundesministerium für Forschung, Technologie und Raumfahrt (BMFTR) via project QuantumHiFi - 16KIS1592K - Forschung Agil. The project is co-financed by ERDF of the European Union and by `Fonds of the Hamburg Ministry of Science, Research, Equalities and Districts (BWFGB)'.
\end{acknowledgments}

\bibliography{references} 

\appendix
\section{Used Components}
\label{sec:appendix}
The following components were used in the imaging system, as labeled in Fig.~\ref{fig:cryo_powerPoint}:\\
\noindent
\textbf{Microscope objective (10):}
EO High Resolution Infinity Corrected Objective, Edmund Optics, Nr.~28-20-45-000, focal length $f_\text{o}=\SI{20}{\milli\meter}$, resolving power (Airy disk radius, comparable to MTF10\%) \SI{0.74}{\micro\meter}, working distance \SI{19}{\milli\meter}. Stable during multiple cooling cycles without any visible damage.\\
\textbf{Relay achromats of upper $4f$ system (6,7):}
Thorlabs: ACT508-250-A - f = 250 mm, Ø2" Achromatic Doublet, ARC: 400 - 700 nm \\
\textbf{Relay achromats of lower $4f$ system (8,9):}
Thorlabs: ACT508-200-A - f = 200 mm, Ø2" Achromatic Doublet, ARC: 400 - 700 nm \\
\textbf{Resolution test target (11):} Thorlabs: R1DS1N1 - Negative 1951 USAF Test Target Groups 0-7, Ø1" \\
\textbf{Nanophotonic diamond sample (11):} LIGHTSYNQ retroreflector sample\\
\textbf{High magnification tube lens (16):} TS-Optics 62 mm f/8.4 4-Element Flatfield Refractor, Quadruplet lens system, $f_\text{t}=\SI{520}{\milli\meter}$, aperture \SI{62}{\milli\meter}\\
\textbf{Telescope camera (16):} ZWO ASI294MC Pro, 19.1 x \SI{13.0}{\milli\meter} sensor, \SI{4.63}{\micro\meter} pixel size\\
\textbf{Digital camera with zoom lens (15):} CANON EOS M50, with lens EF-M 55-200mm, f/4.5-6.3\\
\textbf{Dilution refrigerator:} Bluefors LD250\\
\textbf{White laser (1):} NKT SuperK COMPACT, visible power 20mW, pulse repetition rate \SI{20}{\kilo\hertz}\\
\textbf{Green laser (2):} Thorlabs PL253 - Compact Laser Module with Bare Wire Leads, 520 nm, 4.5 mW \\
\textbf{Galvanometer mirror (3):} Phenixtechnology PT-A40\\
\textbf{Stepper motors for adjusting the telescope and beamsplitter (4, 16):} ZWO Electronic Automatic Focuser (EAF)

\section{Magnification and field of view}
\label{si:magnification}
\begin{figure}[t] 
	\includegraphics[width=\linewidth]{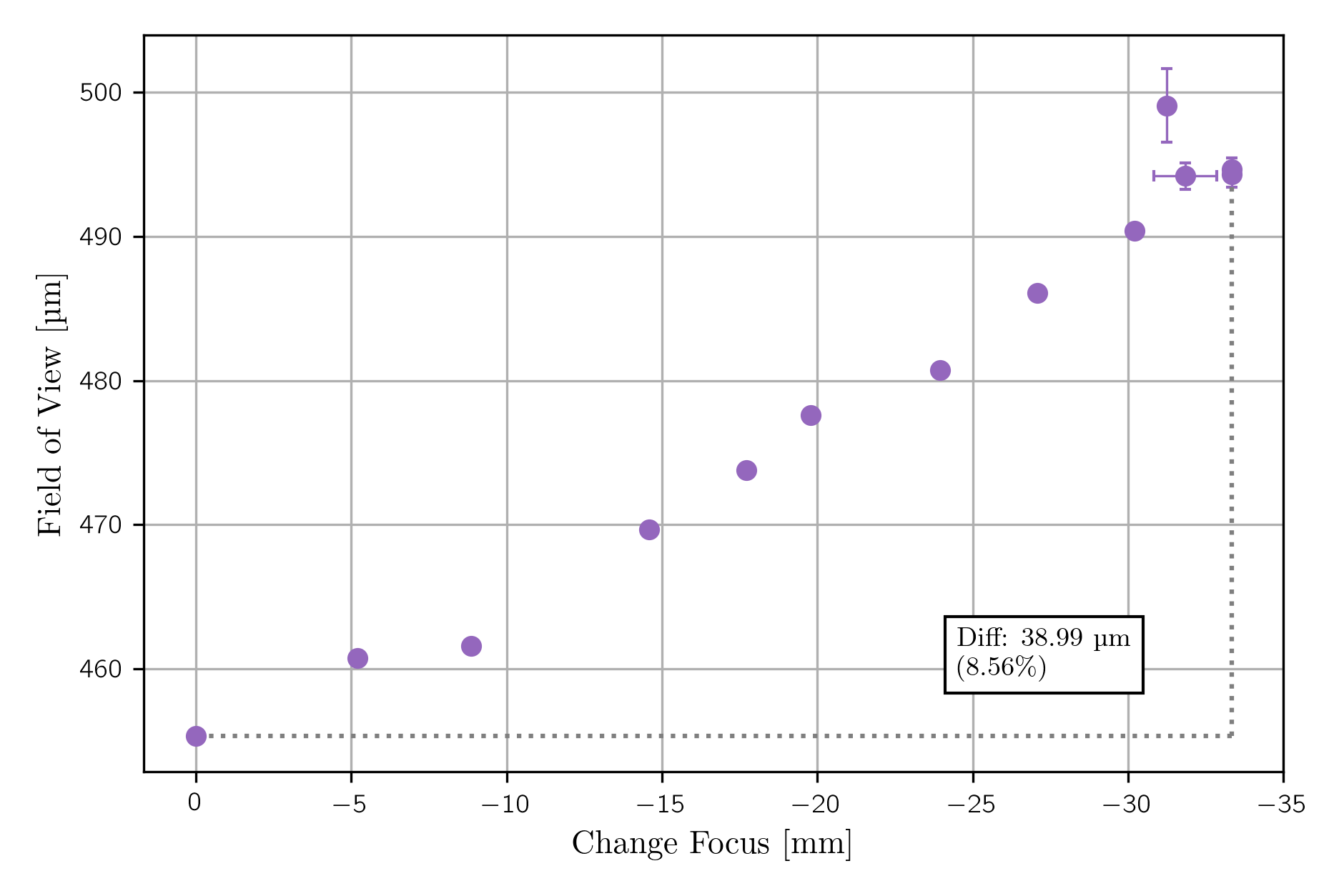}
	\caption{Change of field-of-view measured diagonally caused by focus shift at the telescope camera due to vacuum pumping.}
	\label{fig:diag_pumping}
\end{figure}

Due to changes in focus in the high-magnification setup, i.e.\ at the telescope camera, the magnification and therefore the field-of-view vary. Before closing and evacuating the system, the microscope objective inside the cryostat is adjusted, such that the image plane behind the telescope is sufficiently far away, that the focal shift during evacuation and cool-down can be pre-compensated. Consequently, the effective focal length and hence magnification is substantially increased, compared to the target magnification of $M=26$. 
Based on a sensor diagonal of \SI{23.1}{\milli\meter}, the magnification before evacuation is $M\sim50$, and reduces by approximately 9\% during the pumping process (see Fig.~\ref{fig:diag_pumping}), with an average of $M\sim48$ during evacuation. During cool-down, the image focal plane is shifted closer to the telescope lens, approaching its nominal back focus. This results in a further decrease in magnification (see Fig.~\ref{fig:pump_shift}b) ultimately reaching $M\sim28$, close to the design value of $M=26$. The residual difference in magnification to the design value can be explained by an offset of the telescope's principle plane from the $8f$ point, and due to small offsets in the distance between the relay lenses. Small variations in the final focal plane can occur, with \SI{10}{\milli\meter} shift in the camera position corresponding to merely \SI{12}{\micro\meter} shift on the objective side at a magnification $M\sim28$. Errorbars of focal positions (Fig.~\ref{fig:pump_shift}) are given by statistical standard deviation of the focal position after repeated optimization. 

\begin{figure}[h] 
	\includegraphics[width=\linewidth]{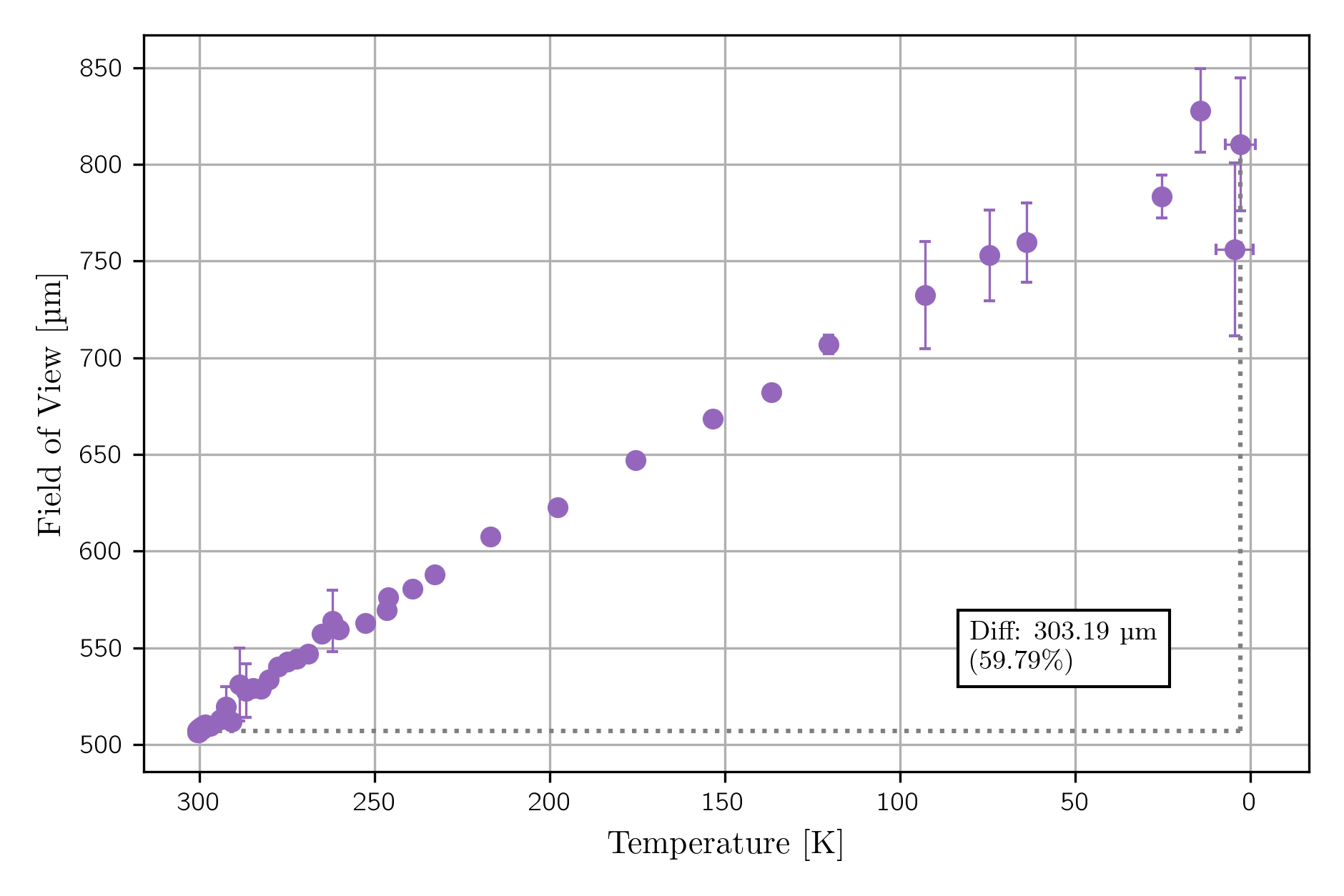}
	\caption{Change of field-of-view measured diagonally caused by focus shift at the telescope camera due to cooling. At the start of the measurement the data collection rate was higher, resulting in more data points above $\SI{250}{\kelvin}$.}
	\label{fig:dia_cool}
\end{figure}

\section{Estimation of focal shift}
\label{si:focal}

\subsection{Focal shift relation}

When the system is perfectly aligned, the sample is on the focal plane of the objective, the relay lenses act as an ideal one-to-one imaging system, and the camera is positioned in the nominal back-focal plane of the telescope. 
Minor misalignment of the $8f$ system has a minimal effect on the rest of the imaging setup: Using the matrix method, one can find that an offset $d_{6,7}$ 
(respectively $d_{8,9}$) between the lenses 6, and 7 in Fig.~\ref{fig:cryo_powerPoint} (resp. 8, 9), results in an effective focal length 
\begin{equation}
    \frac 1 {f_\text{relay}} \approx \frac{d_{6,7}}{f_6^2} + \frac{d_{8,9}}{f_8^2}. 
\end{equation}
Due to the long focal lengths and small offsets, this is negligible compared to the relatively short focal lengths of the telescope and objective. 
An offset $d_{7,8}$ between lenses 7 and 8, simply appears as an additional offset between the $8f$ points and the principle planes of the telescope $d_{6,16}$ and objective $d_{9,10}$. 

The entire lens system can be modeled as a compound thick lens, with the effective focal length approximated by
\begin{equation}
    \frac{1}{f_\text{eff}} \approx \frac{1}{f_\text{o}}+\frac{1}{f_\text{t}} - \frac{d}{f_\text{o} f_\text{t}},
\end{equation}
where $d\approx d_{7,8}+d_{6,16}+d_{9,10}$ is the sum of offsets of principle planes. 
Once the camera and objective are aligned to form a focused image, we quantify the distance of the sample from the objective-side principle plane of the compound system $s_\text{o}$ and the of the camera from the telescope-side principle plane $s_\text{t}$, and such that we can describe the image distance relations as
\begin{equation}
    \frac{1}{f_\text{eff}} = \frac{1}{s_\text{o}}+\frac{1}{s_\text{t}}.
\end{equation}
The magnification of the compound system is 
\begin{equation}
    M=\frac{s_\text{t}}{s_\text{o}}.
\end{equation}
When the system is perfectly aligned, i.e.\ the back principle plane of the objective and the front principle plane of the telescope overlap exactly with the respective $8f$ points, we find $s_\text{o}=f_\text{o}$ and $s_\text{t}=f_\text{t}$, yielding the nominal magnification $M_\text{nom}=f_\text{t}/f_\text{o}$.  

Next, we investigate how the system behaves under small shifts, $s_\text{o}\rightarrow s_\text{o}+\delta s_\text{obj}$, $s_\text{t}\rightarrow s_\text{t}+\delta s_\text{tel}$, and $f_\text{eff}\rightarrow f_\text{eff}+\Delta f$, such that the lens equation becomes
\begin{equation}
    \frac{1}{f_\text{eff}+\delta f} = \frac{1}{s_\text{o}+\delta s_\text{obj}}+\frac{1}{s_\text{t}+\delta s_\text{tel}}.
\end{equation}
Assuming that the change in focal length is small, we can solve for small
$\delta s_\text{t}$
\begin{equation}
   \delta s_\text{tel} \approx -\frac{ \left(\delta s_\text{obj}-\delta f(1+1/M^2)\right) M^2}{1+\frac{\left(\delta s_\text{obj}-\delta f(1+1/M^2) \right)}{s_\text{obj}} (M+1)},
\end{equation}
with the magnification evaluated before the shift. 
For large magnification, we can conclude that a change in the effective focal length can be treated like an opposite change in the sample position. 

We can thus approximate the relative focal shift as
\begin{equation}
    \frac{\partial s_\text{t}}{\partial s_\text{o}}= -M^2. 
\end{equation}
and 
\begin{equation}
    \frac{\partial s_\text{t}}{\partial f_\text{eff}} = M^2+1. 
\end{equation}
Note that the magnification can change substantially for larger shifts of the objective. As the position of the principle planes of the objective and telescope are not well known, we extract the derived values based on the measured magnification: for the change in focal length due to the reduction in pressure, we find
\begin{equation}
    \Delta f = \int \frac{\text{d}s_\text{t}}{M^2(s_\text{t})+1},
\end{equation}
with the magnification in Fig.~\ref{fig:diag_pumping}, and the focal shift in Fig.~\ref{fig:pump_shift}. This shift in focal length can be treated as an opposite shift in the image focal plane $\Delta s_\text{obj}\approx -\Delta f$. 
For thermal contraction, we find the change between the front principle plane and the device  
\begin{equation}
    \Delta s_\text{obj} = -\int \frac{\text{d}s_\text{t}}{M^2(s_\text{t})}
\end{equation}
using the temperature dependent data in Fig.~\ref{fig:dia_cool} and \ref{fig:pump_shift}b. As the absolute value of $s_\text{t}$ is not known, we numerically evaluate these integrals over relative shifts of the image focal plane $\Delta s_\text{tel}=s_\text{t}-\text{const}$.

We note that the magnification changes more than intuitively expecting, indicating that the principle plane of the objective is substantially offset for the $8f$ point. This is not surprising, as the principle planes of the objective are unknown. The offset is small enough such that it does not negatively affect the system performance. 

\subsection{Thermal contraction}
The objective has a length of $L_\text{O}=\SI{76}{\milli\meter}$ and is mounted to a copper holder, which in turn is mounted to copper side rails. The sample is in focus, i.e.\ separated from the objective holder by the parfocal length of $L_\text{C}=\SI{95}{\milli\meter}$, exclusively by copper components. Hence, the distance between the front lens and the sample changes by 
\begin{equation}
    \Delta z_\text{therm} =  -\alpha_\text{c, Cu} * L_\text{C} + -\alpha_\text{c, O} * L_\text{O} ,
\end{equation}
with the literature value for the cumulative thermal contraction $\alpha_\text{c, Cu}=\Delta L_\text{C}/L_\text{C} = 0.324\%$ for copper and $\alpha_\text{c, O}$ for the objective, which is unknown. As the objective is non-magnetic, the usual materials used in objectives are aluminum ($\alpha_\text{c, Al} = 0.415\%$), brass ($\alpha_\text{c, CuPbZn} = 0.384\%$) and polymers, or any combination thereof. X-ray tomography of the objective reveals structures with heavier nuclei (consistent with brass), lighter nuclei (consistent with aluminum), and some components that are not visible (consistent with carbon-based materials) around the front lens. The image quality is not sufficient to construct a reasonable model of the objective, but supports the hypothesis that some components are held in place by a polymer. 

\subsection{Focal length changes}

Using the lensmaker's equation in the thin-lens approximation, we estimate the effect of refractive index change of air due to evacuation, and of the lens material due to the thermo-optic effect. As the front lens of the objective has a high NA, the thin lens approximation is not accurate, but gives a viable first estimate. Corrections for lens thickness are small compared to the values obtained by this approach. 
Equation \eqref{eq:f_change_evacuation} is obtained by using the pressure dependent refractive index of air $n_\text{air} = 1.0003 + \delta n_\text{air}$, and linearizing the lensmaker's equation in $\delta n_\text{air}$: 
\begin{equation}
    f' = \frac{f}{1 - \delta n_\text{air} (1+1/(n_\text{lens}-1))}, 
\end{equation}
Vacuum is reached for a differential pressure $\delta p=-\SI{1013}{\hecto\pascal}$. Hence, we find the focal shift of $\Delta f = f'-f\approx -7.3\cdot 10^{-4} \cdot f_o$
for a high refractive index lens ($n_\text{lens}=1.7$), and $\Delta f = f'-f\approx -9.7\cdot 10^{-4} \cdot f_\text{o}$ for a low index lens ($n_\text{lens}=1.45$). For the $10\times$ objective used here, this corresponds to shortening of the focal length by between \SI{14.6}{\micro\meter} and \SI{19.3}{\micro\meter}, matching well with the observed \SI{14.4}{\micro\meter}.
We note that this change only effects curved surfaces. As the front lens of an objective is approximately plano-convex, the sample-facing surface is contributing minimally to the overall change in focal length. This indicates, that the inside of the objective is evacuated well. 

For the thermo-optic effect \eqref{eq:f_change_therm}, we use the absolute refractive index of a single lens $n_\text{lens,T}= n_\text{lens} + \delta n_\text{lens}$, and linearize the lensmaker's equation in $\delta n_\text{lens}$:
\begin{equation}
    f' = \frac{f}{1 + \delta n_\text{lens}/ (n_\text{lens}-1)}.
\end{equation}

Hence the focal shift $f'-f\approx -\delta n_\text{lens}/(n_\text{lens} -1) \cdot f_o$. In contrast to thermal contraction, the thermo-optic effect is strongly nonlinear in temperature, in various cases of common glasses even reversing its sign. As the focal change appears approximately linear in temperature, this does not appear to be a major contributor to the overall effect. 

Similarly, the focal length of a lens changes with thermal contraction of the lens itself. The thermal expansion of glass varies from material to material, but is generally less than that of most metals. Extrapolating the cumulative relative thermal contraction $\alpha_\text{c, lens}$ from the room-temperature coefficient of linear expansion as a first order approximation, we find the focal shift $f'-f\approx - \alpha_\text{c, lens} \cdot f$. Observing that N-BK7 and N-SF2 have a relatively large coefficient of linear thermal expansion, which is similar to that of titanium, we approximate $\alpha_\text{c, lens}  \lesssim \alpha_\text{c, Ti} = 0.15\%$.

\FloatBarrier

\end{document}